\documentclass[epj,english,twocolumn]{webofc}
\usepackage[varg]{txfonts}   
\usepackage{hyperref}        
\usepackage{multicol}
\usepackage{graphicx}
\usepackage{amsmath}
\woctitle{Heavy Ion Accelerator Symposium 2013}
\begin{document}
\title{Clusters, Halos, And S-Factors In Fermionic Molecular Dynamics
   \thanks{Supported by the ExtreMe Matter Institute EMMI} }

\author{Hans Feldmeier \inst{1} \and  Thomas Neff  \inst{1} }
\institute{GSI Helmholtzzentrum f{\"u}r Schwerionenforschung GmbH, Darmstadt}

\abstract{
In Fermionic Molecular Dynamics antisymmetrized products of Gaussian wave packets 
are projected on angular momentum, linear momentum, and parity. An appropriately 
chosen set of these states span the many-body Hilbert space in which the Hamiltonian is diagonalized.  
The wave packet parameters  -- position, momentum, width and spin --  are obtained by variation under 
constraints.
The great flexibility of this basis allows to describe not only shell-model like states but also exotic states
like halos, e.g. the two-proton halo in $^{17}$Ne, or cluster states as they appear for example in $^{12}$C 
close to the $\alpha$ breakup threshold where the Hoyle state is located.
Even a fully microscopic calculation of the $^3$He($\alpha$,$\gamma$)$^7$Be capture reaction 
is possible and yields an astrophysical S-factor that compares very well with newer data.
As representatives of numerous results these cases will be discussed in this contribution, 
some of them not published so far.
The Hamiltonian is based on the realistic Argonne~V18 nucleon-nucleon interaction.
          }


\newcommand{\bra}[1]{\big< \,{#1}\, \big| }
\newcommand{\ket}[1]{\big| \,{#1}\, \big> }
\newcommand{\braket}[2]{\big\langle {#1} \big| {#2} \big\rangle}
\newcommand{\matrixe}[3]{\big< \,{#1}\, \big| \,{#2}\, 
  \big| \,{#3}\, \big> }

\renewcommand{\vec}[1]{\mathbf{#1}}
\newcommand{\op}[1]{\hat{#1}}

\maketitle


\section{Fermionic Molecular Dynamics (FMD)}

In the FMD approach we employ Gaussian wave packets
\begin{equation}
  \label{eq:wavepacket}
  \braket{\vec{x}}{q} = \exp\left\{-\frac{(\vec{x}-\vec{b})^2}{2 a}\right\}
   \otimes \ket{\chi^\uparrow, \chi^\downarrow} \otimes \ket{\xi}
\end{equation}
as single-particle basis states. The complex parameters $\vec{b}$ encode the mean positions and
momenta of the wave packets and $a$ the widths of the wave packets. The spins can
assume any direction, isospin is $\pm 1$ denoting a proton or a
neutron. Intrinsic many-body basis states are Slater determinants 
\begin{equation}
  \label{eq:sldet}
  \ket{Q} = \mathcal{A} \left\{\ \ket{q_1} \otimes \ldots \otimes \ket{q_A}\ \right\} \: 
\end{equation}
that reflect deformation or clustering and break the symmetries 
of the Hamiltonian with respect to parity, rotation and translation. To restore the symmetries 
the intrinsic basis states are projected on parity, angular momentum and total linear momentum
\begin{equation}
 \ket{Q; J^\pi MK; \vec{P}=0} = \op{P}^\pi \op{P}^J_{MK} \op{P}^{\vec{P}=0} \; \ket{Q} .
\end{equation}
In a full FMD calculation the many-body Hilbert space is spanned by a set of $N$ 
projected intrinsic basis states 
$\left\{\ \ket{Q^{(a)}; J^\pi MK; \vec{P}=0} , a=1,\ldots,N \ \right\}$. 
By diagonalizing the Hamiltonian in this set of non-orthogonal basis states
the amplitudes of the various configurations contained in the many-body eigenstate 
are determined.

Starting from the realistic Argonne~V18 interaction \cite{wiringa95}
we derive a phase-shift-equivalent effective low-momentum interaction using the unitary
correlation operator method (UCOM). 
The basic idea of the UCOM
approach is to explicitly include short-range central and tensor
correlations by means of a unitary operator
\cite{ucom98,ucom03,ucom10}. 
No-core shell model calculations show that the two-body UCOM interaction gives
a good description of $s$- and light $p$-shell nuclei \cite{ucom10},
indicating that the neglected induced 3-body forces cancel to a certain extent 
the missing genuine 3-body forces.

\section{Cluster States in $^{12}$C}

The structure of the second $0^+_2$ state in $^{12}$C, the Hoyle state, 
is enjoying renewed and still growing interest in nuclear structure research 
\cite{freer12,zimmermann13,epelbaum11,epelbaum12,neff12}. 
In \cite{Hoyle07} we investigated its structure with a model space spanned by
angular momentum projected FMD configurations obtained by variation 
plus a full set of projected three-$\alpha$ triangular configurations. 
We found that the Hoyle state
is very dilute and extended, consisting mainly of well distinguished $\alpha$-clusters.  
This is illustrated in the top part of Fig.~\ref{fig:intrinsic} where we show the density 
distribution of those intrinsic FMD basis states that have the largest overlap with the 
ground state and the Hoyle state.

%
 \newcommand{\bk}[2]{\Big|\Big<\,#1 \;\Big|\;#2 \,\Big>\Big|^2}
\begin{figure*}[!tb]
\centering
\begin{tabular}{c|cccc}
\hline
  \includegraphics[width=0.15\textwidth]{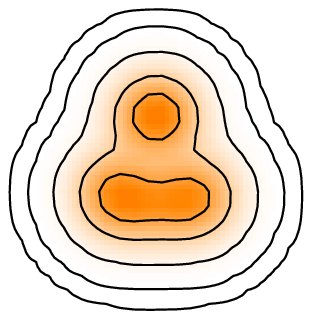}&
  \includegraphics[width=0.15\textwidth]{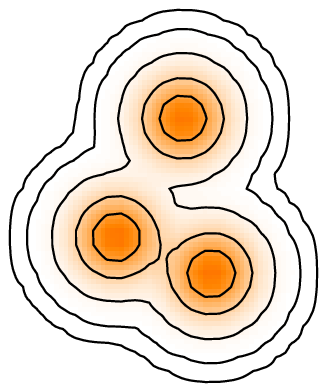}&
  \includegraphics[width=0.15\textwidth]{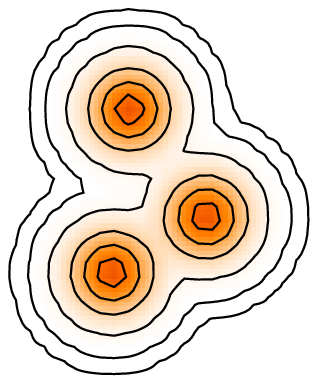}&
  \includegraphics[width=0.15\textwidth]{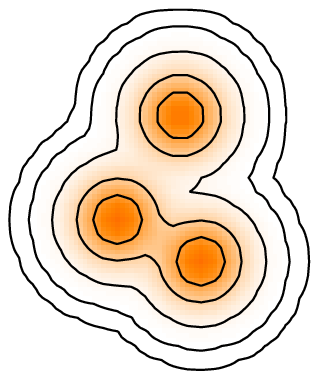}&
  \includegraphics[width=0.15\textwidth]{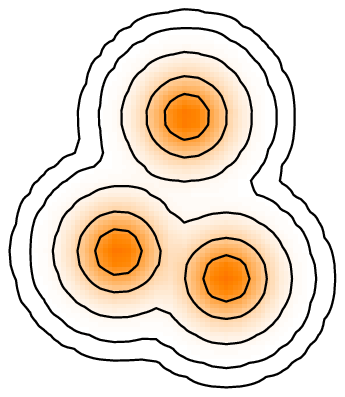}\\[-7mm]
  \parbox{0.15\textwidth}{\center\footnotesize$\bk{\cdot}{0_1^+}=0.88$}& 
  \parbox{0.15\textwidth}{\center\footnotesize$\bk{\cdot}{0_2^+}=0.54$}&
  \parbox{0.15\textwidth}{\center\footnotesize$\bk{\cdot}{0_2^+}=0.52$}&
  \parbox{0.15\textwidth}{\center\footnotesize$\bk{\cdot}{0_2^+}=0.52$}&
  \parbox{0.15\textwidth}{\center\footnotesize$\bk{\cdot}{0_2^+}=0.44$}\\[3mm]
\hline
  \includegraphics[width=0.15\textwidth]{C12-0_1-1.fmd.dens.eps}&
  \includegraphics[width=0.15\textwidth]{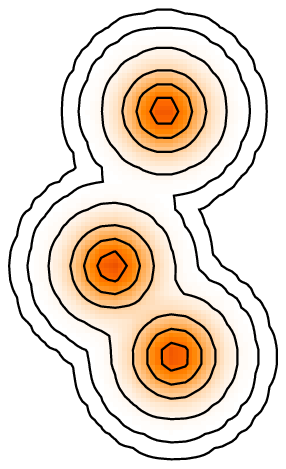}&
  \includegraphics[width=0.15\textwidth]{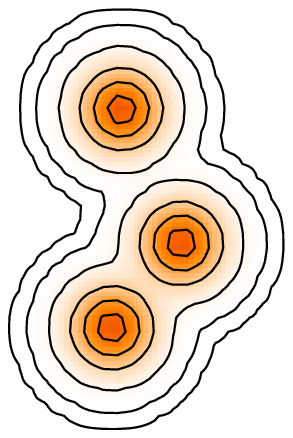}&
  \includegraphics[width=0.15\textwidth]{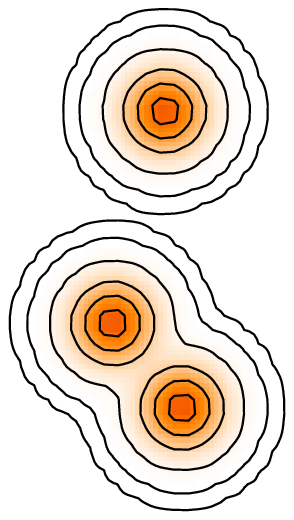}&
  \includegraphics[width=0.15\textwidth]{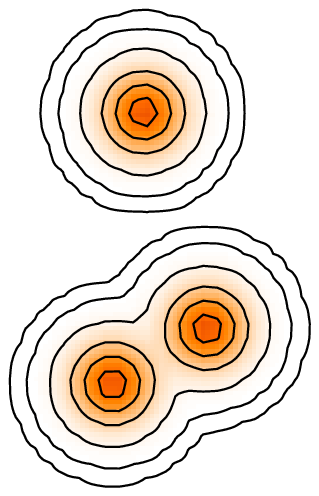}\\[-7mm]
  \parbox{0.15\textwidth}{\center\footnotesize$\bk{\cdot}{2_1^+}=0.86$}&
  \parbox{0.15\textwidth}{\center\footnotesize$\bk{\cdot}{2_2^+}=0.34$}&
  \parbox{0.15\textwidth}{\center\footnotesize$\bk{\cdot}{2_2^+}=0.30$}&
  \parbox{0.15\textwidth}{\center\footnotesize$\bk{\cdot}{2_2^+}=0.23$}&
  \parbox{0.15\textwidth}{\center\footnotesize$\bk{\cdot}{2_2^+}=0.21$}\\[3mm]
\hline
\end{tabular}
 \caption{Top: density of intrinsic FMD basis states that have the largest overlaps with 
  the ground ($0^+_1$)   and Hoyle ($0^+_2$) state, respectively. Note that FMD states are 
  not orthogonal. 
   Bottom: same as top, but for the for $2^+_1$ and $2^+_2$ states. 
  The $0^+_1$ and the  $0^+_2$ states have different intrinsic structures and thus the
  recently identified $0^+_2$ is not just simply a rotating Hoyle state. 
 \label{fig:intrinsic}}
\end{figure*}
\begin{figure}[hhh]
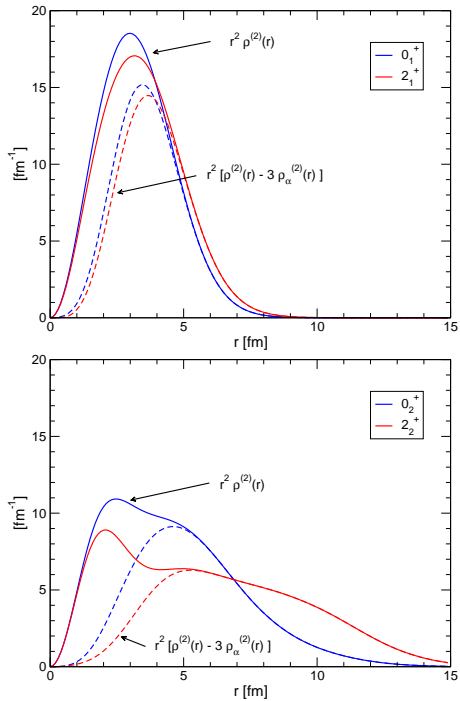

\centering
  \includegraphics[width=0.35\textwidth]{C12-tbdens-gs-band.eps}\\
  \includegraphics[width=0.35\textwidth]{C12-tbdens-hoyle-band.eps}
 \caption{Top: full lines denote two-body densities as function of distance r between particle pairs 
of $^{12}$C ground state rotational band members  $0^+_1$(blue) and  $2^+_1$ state (red); 
dashed lines show result when the distribution of pairs where both particles are inside the same 
$\alpha$-cluster are subtracted \cite{neff12}. This indicates distance distributions of pairs, where 
the two particles are in different $\alpha$-clusters. Bottom: same as top, but for the Hoyle state  
"rotational band" members $0^+_2$(blue) and  $2^+_2$ state (red). See also Fig.~\ref{fig:intrinsic}.  
 \label{fig:two-body}
 }
\end{figure}
While the leading intrinsic configuration of the ground state is very compact, and after projection on
good angular momentum, essentially a shell model state filling the $p_{3/2}$-shell,
the Hoyle state is a quantal superposition of three $\alpha$-clusters arranged in a
slightly opened triangle configuration, or one may regard it also as a $^8$Be surrounded by an $\alpha$-cluster,
see upper part of Fig.~\ref{fig:intrinsic}.

The first $2^+$ state has the same leading intrinsic configuration as the ground state 
and may thus be regarded as the $J^\pi=2^+$ member of a rotational band based on the ground state. 
The analogue argument does not hold for the second $2^+$ state, it does not quite look like
the $2^+$ member of a rotational band with the same intrinsic structure as the Hoyle state
(lower part of Fig.~\ref{fig:intrinsic}). 
There is still the  $^8$Be correlation but the third $\alpha$-cluster   
is pushed further away by centrifugal forces and feels attraction to only one of the
$^8$Be $\alpha$-clusters, thus forming obtuse triangles.

Although intrinsic FMD basis states provide an intuitive understanding of the structure of the many-body
state, they are not observable. Therefore we proposed in Ref.~\cite{neff12} to look at the two-body density 
\begin{equation}
\rho^{(2)}(\vec{r})=\matrixe{\Psi}{\sum_{i<j}\delta(\op{\vec{r}}_i-\op{\vec{r}}_j-\vec{r})}{\Psi}\ ,
\end{equation}
which gives the probability to find a pair of nucleons at a distance $\vec{r}$. This 
correlation function can be calculated in any representation and can thus be used to compare
different many-body approaches.

Fig.~\ref{fig:two-body} shows that in the $2^+_1$ state the distances between nucleons are slightly larger
than in the ground state but otherwise very similar. The lower part of  Fig.~\ref{fig:two-body}
reveals  much larger particle distances for the  $0^+_2$ (Hoyle) and the $2^+_2$ state.
In the Hoyle state a shoulder appears around $5$~fm indicating the pairs where one particle is in one
$\alpha$-cluster and the other one in the neighbouring $\alpha$-cluster, compare Fig.~\ref{fig:intrinsic} 
($5$~fm is the typical distance between the centers of the clusters). 
The maximum around $2$~fm originates from pairs within the same cluster. The picture becomes even more 
transparent when we subtract three times the pair distributions of a single 
$\alpha$-cluster (dashed lines). The distribution of the dilute Hoyle-like $2^+_2$ state
shows a broad shoulder between 5 and 10~fm coming from pairs with one nucleon in the
distant $\alpha$-cluster and one in the $^8$Be-like structure, see lower part of Fig.~\ref{fig:intrinsic}.

Both, the recently measured energy of the $2^+_2$ resonance and its $B(E2)$-value of the transition 
to the ground state \cite{zimmermann13} compare well with our predictions, see Table~\ref{table:12C}.

We used these many-body wave functions also to calculate the transition form factor 
from the ground state to the Hoyle state and compared it directly to electron scattering
data \cite{Hoyle07,Hoyle10}. 
Our results are similar to those of \cite{kamimura81, funaki03} where a many-body state 
representing a gas of independent $\alpha$-clusters is assumed.
The good agreement between calculated and measured form factors is a strong
confirmation for a spatially extended cluster structure of the Hoyle state. 
The overall agreement with many other measured data indicates that the FMD description 
gives a good insight into the structure of $^{12}$C.

\begin{table}[!h]
\centering \caption{Radii and transitions in $^{12}$C,
            data: \cite{ajzenbergselove90,itoh04,zimmermann13}\label{table:12C}}
\begin{tabular}{lrr}
\hline\\[-3mm]
Energies [MeV]                   &    Exp~~   &    FMD~     \\
\hline\\[-3mm] 
$E(0_1^+)$                       & -92.16     & -92.64         \\[1mm]
$E^*(2_1^+)$                     &   4.44     &   5.31         \\[1mm]
$E(3\alpha)$                     & -84.89     & -83.59         \\[1mm]
$E(0_2^+)-E(3\alpha)$            &   0.38     &   0.43         \\[1mm]
$E(2_2^+)-E(3\alpha)$            &2.76(11)    &   2.77         \\
\hline\\[-3mm]
          Radii [fm]            &    Exp~~    &    FMD~     \\
\hline\\[-3mm] 
$r_\mathrm{charge}(0_1^+)$       & 2.47(2)    & 2.53        \\[1mm]
$r(0_1^+)$                       &            & 2.39        \\[1mm]
$r(2_1^+)$                       &            & 2.50        \\[1mm]
$r(0_2^+)$ Hoyle state           &            & {\bf 3.38}  \\[1mm]
$r(2_2^+)$ Hoyle like            &            & {\bf 4.43}  \\[1mm]
\hline\\[-3mm]
Transitions  [fm$^2$] or [e$^2$fm$^4$] &     Exp~~    &  FMD~    \\
\hline\\[-3mm] 
$M(E0, 0_1^+\!\rightarrow 0_2^+$)      & 5.4(2)        & 6.53     \\[1mm]
$B(E2, 2_1^+\!\rightarrow 0_1^+$)      & 7.6(4)        & 8.69     \\[1mm] 
$B(E2, 2_1^+\!\rightarrow 0_2^+$)      & 2.6(4)        & 3.83     \\[1mm]
$B(E2, 2_2^+\!\rightarrow 0_1^+$)      & 0.73(13)      & 0.46     \\[1mm]
\hline
\end{tabular}
\end{table}

Ab initio nuclear lattice calculations \cite{epelbaum11,epelbaum12} seem to support this structure 
but due to the large lattice constant the angles and sites of the three-$\alpha$ triangles can 
assume only discrete values in the sampled configurations. 
For example the typical distance between the $\alpha$-clusters in the Hoyle state
is 2 to 3 lattice spacings.  
It will be very interesting to see if future more refined calculations on the nuclear lattice
will confirm further the FMD results on the numerous aspects of the $^{12}$C structure.

\section{Neon isotopes and two-proton halo}
\begin{figure}[h]
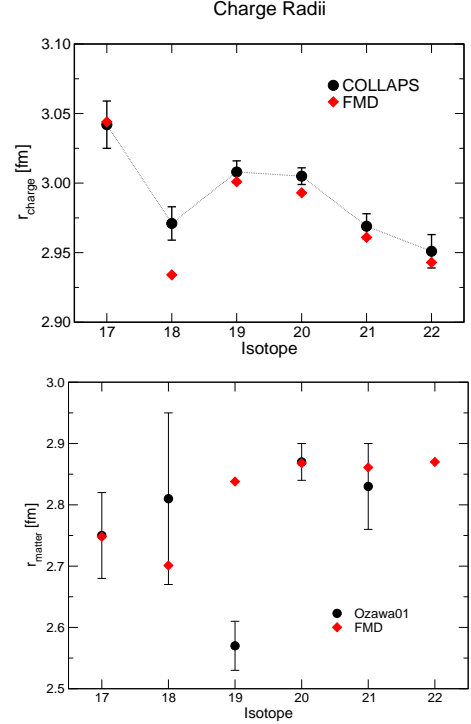

\center
  \includegraphics[width=0.35\textwidth]{Neon-charge-radii.eps}\\[2mm]
 ~~\includegraphics[width=0.345\textwidth]{Neon-matter-radii-hfmd+cluster.eps}
 \caption{Top: charge radii of Ne isotopes measured by COLLAPS und calculated with FMD
 \cite{geithner08}
   Bottom: point mass radii, data from \cite{ozawa01}   }
 \label{fig:neon-radii}
\end{figure}
%
\begin{figure}[h]
\center 
  \includegraphics[width=0.4\textwidth]{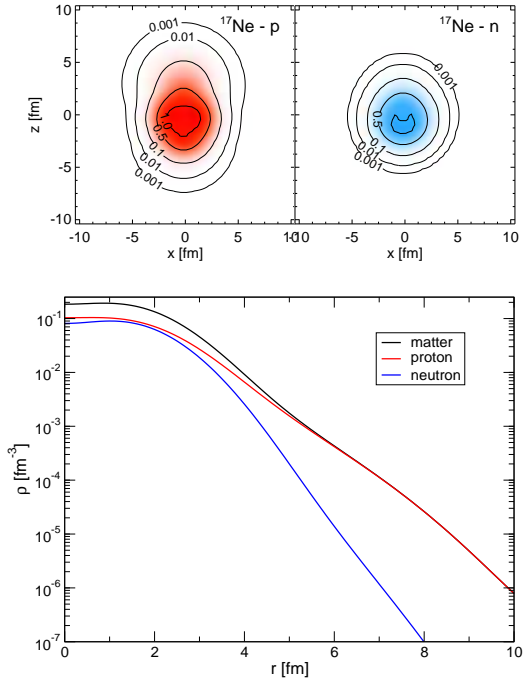}
  \includegraphics[width=0.4\textwidth]{Ne17-densities-log.eps}
 
 \caption{Top: proton and neutron density of leading intrinsic FMD state contributing
to ground state of $^{17}$Ne, bottom: charge distribution calculated with FMD 
eigenstate showing a two-proton  halo \cite{geithner08}.
\label{fig:neon-halo}          }
 
\end{figure}
The charge radii of the neon isotopes, which have been measured in 
Ref.~\cite{geithner08}, do not show the usual monotonic increase with
mass number, while the matter radii seem to increase
monotonically from $^{18}$Ne on, see Fig.~\ref{fig:neon-radii}. 
The FMD model explains this by substantial changes in the ground-state
structure. It attributes the large charge radius of $^{17}$Ne to an extended 
two-proton halo, as seen in Fig.~\ref{fig:neon-halo}, with an $s^2$ component of about 40\%.
The leading intrinsic state (upper part of Fig.~\ref{fig:neon-halo}) has a far out reaching part 
consisting of two correlated protons, while the neutron distribution is almost spherical 
and only weakly polarized by the outer protons. This is in accord with the simplest picture that
$^{17}$Ne consists of an $^{15}$O core plus two protons in either $s^2$ or $d^2$ configurations.
Interaction cross sections \cite{ozawa94}
and longitudinal momentum distributions \cite{kanungo05} support the halo picture.
In $^{18}$Ne the situation is similar but in the same simple picture the core is now the doubly magic
$^{16}$O, which leads to a significantly smaller $s^2$ component and hence a smaller charge radius.
 
The subsequent increase in charge radius  for $^{19}$Ne is of different origin. 
The fact that the experimental $1/2^+$ and $1/2^-$ states are almost degenerate 
and the cluster thresholds are pretty low, hints already at possible admixtures of 
states with cluster structures.
Therefore it is not surprising that $^{16}$O~-~$^3$He and $^{15}$O~-~$^4$He cluster configurations 
admix in the tail of the wave function. 

This admixture of cluster configurations is still very strong in $^{20}$Ne but becomes smaller 
for heavier Ne isotopes explaining the dropping charge radii. 

\begin{figure}[tt]
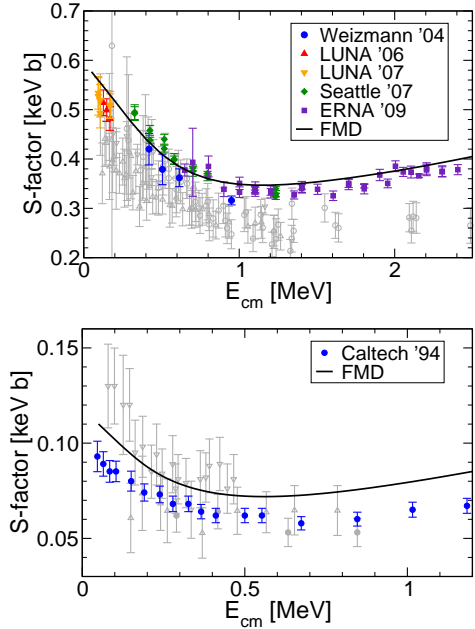

  \centering
  ~~\includegraphics[width=0.35\textwidth]{sfactor-he3-alpha-gamma-be7.eps}\\[2mm]
  \includegraphics[width=0.36\textwidth]{sfactor-h3-alpha-gamma-li7.eps}
  \caption{$S$-factor for capture reaction - top: $^3$He($\alpha$,$\gamma$)$^7$Be,
     recent data
    \cite{narasingh04,bemmerer06,confortola07,brown07,dileva09} colored symbols, 
     older data  gray symbols - bottom:
     $^3$H($\alpha$,$\gamma$)$^7$Li, 
    recent data \cite{brune94} colored, older data gray symbols.}
  \label{fig:sfactor}
\end{figure}

\section{Radiative capture reaction $\mathbf{^3\mathrm{He}(\alpha,\gamma)^7\mathrm{Be}}$ }

Another application of the FMD approach is the calculation of the 
$^3$He($\alpha$,$\gamma$)$^7$Be radiative capture reaction \cite{neff11}. As this reaction 
plays an important role in the solar proton-proton chains and determines the
production of $^7$Be and $^8$B neutrinos \cite{adelberger98,adelberger11}, it 
has been studied extensively from the experimental side in recent years
\cite{narasingh04,bemmerer06,confortola07,brown07,dileva09}.  However,
it is still not possible to reach the low energies relevant for solar burning in experiment.
From the theory side this reaction has been investigated using simple
potential models, where $^3$He and $^4$He are treated as point-like
particles interacting via an effective nucleus-nucleus potential,
e.g., \cite{kim81} or microscopic cluster models, e.g.,
\cite{langanke86,kajino86} where the $^7$Be bound and scattering
states are constructed from microscopic $^3$He and $^4$He clusters
interacting via an effective nucleon-nucleon
interaction. \emph{Ab-initio} calculations using variational Monte
Carlo \cite{nollett01} and no-core shell model wave functions
\cite{navratil07} were used to calculate asymptotic normalization
coefficients for the bound states but relied on potential models for the scattering 
phase shifts. 

In the FMD calculation we divided the many-body Hilbert space
into an external region, where the scattering states are antisymmetrized products of 
$^3$He and $^4$He clusters in their FMD ground states at various distances, and 
an interaction region, where FMD configurations were obtained by variation after
projection on spin-parity $1/2^+$, $3/2^+$,
$5/2^+$ and $3/2^-$, $1/2^-$, $7/2^-$, $5/2^-$. A constraint on the radius of the intrinsic
states was used to vary the distance between the clusters. 
Using the microscopic $R$-matrix  method \cite{descouvemont10}
boundary conditions for bound and scattering states were implemented by
matching to Whittaker and Coulomb functions at the channel radius ($a$ = 12~fm). 

The capture cross section was calculated from electromagnetic transition rates between the 
microscopic many-body scattering and bound states. The result for the total cross section 
for the $^3$He($\alpha$,$\gamma$)$^7$Be 
capture is shown in form of the astrophysical $S$-factor in the upper part of 
Fig.~\ref{fig:sfactor}. It agrees very well with the recent
experimental data, both in absolute normalization and in the energy dependence. 
The results for the isospin mirror reaction $^3$H($\alpha$,$\gamma$)$^7$Li 
is shown in the lower part of Fig.~\ref{fig:sfactor}. Whereas the energy dependence of the
calculated $S$-factor agrees well with the data, the absolute
cross section is larger then the data by Brune \textit{et al.} by
about 15\%. This is surprising as the FMD results for the $^7$Li bound states and the scattering 
phase shifts are of similar quality as those for $^7$Be.




\bibliographystyle{aipproc}
\bibliography{all}
\IfFileExists{\jobname.bbl}{}
 {\typeout{}
  \typeout{******************************************}
  \typeout{** Please run "bibtex \jobname" to obtain}
  \typeout{** the bibliography and then re-run LaTeX}
  \typeout{** twice to fix the references!}
  \typeout{******************************************}
  \typeout{}
 }

\end{document}